\documentclass[journal]{IEEEtran}
\usepackage{cite}
\usepackage{ifpdf}
\usepackage{amsmath}
\usepackage{amsfonts}
\usepackage{array}
\usepackage{graphicx,bm}
\usepackage{graphicx}
\usepackage{enumitem}
\usepackage{cleveref}

\renewcommand{\b}{\mathbf{b}}

\newcommand{\f}{\mathbf{f}}

\renewcommand{\L}{\mathbf{L}}

\newcommand{\thet}{\bm{\theta}}

\newcommand{\y}{\mathbf{y}}
\renewcommand{\H }{\mathbf{H}}
\newcommand{\z}{\mathbf{z}}
\newcommand{\w}{\mathbf{w}}

\newcommand{\W}{\mathbf{W}}

\newcommand{\0}{\mathbf{0}}

\newcommand{\x}{\mathbf{x}}

\newcommand{\M}{\mathbf{M}}

\renewcommand{\u}{\mathbf{u}}
\renewcommand{\v}{\mathbf{v}}

\newcommand{\q}{\mathbf{q}}

\renewcommand{\b}{\mathbf{b}}

\newcommand{\EE}[1]{{\rm{E}}\left\{#1\right\}}

\renewcommand{\log}[1]{{\rm{log}}#1}

\newcommand{\RNum}[1]{\uppercase\expandafter{\romannumeral #1\relax}}

\begin{document}

\title{Learning to Detect}

\author{Neev~Samuel,~\IEEEmembership{Member,~IEEE,}
        and~Tzvi~Diskin,~\IEEEmembership{Member,~IEEE}
        and~Ami~Wiesel,~\IEEEmembership{Member,~IEEE}
\thanks{Manuscript received May, 2018;
}
\thanks{N. Samuel, T. Diskin and A. Wiesel are with the School of Computer Science and Engineering, The Hebrew University of Jerusalem, Israel.
E-mail: (neev.samuel@mail.huji.ac.il or see http://www.cs.huji.ac.il/~amiw/).}

\thanks{This research was partly supported by the Heron Consortium and by ISF grant 1339/15.}
}
\maketitle

\begin{abstract}
In this paper we consider Multiple-Input-Multiple-Output (MIMO) detection using deep neural networks. We introduce two different deep architectures: a standard fully connected multi-layer network, and a Detection Network (DetNet) which is specifically designed for the task. The structure of DetNet is obtained by unfolding the iterations of a projected gradient descent algorithm into a network. We compare the accuracy and runtime complexity of the purposed approaches and  achieve state-of-the-art performance while maintaining low computational requirements. Furthermore, we manage to train a single network to detect over an entire distribution of channels.
Finally, we consider detection with soft outputs and show that the networks can easily be modified to produce soft decisions.
\end{abstract}

\begin{IEEEkeywords}
MIMO Detection, Deep Learning, Neural Networks.
\end{IEEEkeywords}

%
\IEEEpeerreviewmaketitle

\section{Introduction}
%
%
%
%
\IEEEPARstart{M}{ultiple} input multiple output (MIMO) systems enable enhanced performance in communication systems, by using many dimensions that account for time and frequency resources, multiple users, multiple antennas and other resources. While improving performance, these systems present difficult computational challenges when it comes to detection since the detection problem is NP-Complete, and there is a growing need for sub-optimal solutions with polynomial complexity.

Recent advances in the field of machine learning, specifically the success of deep neural networks in solving many problems in almost any field of engineering, suggest that a data driven approach for detection using machine learning may present a computationally efficient way to achieve near optimal detection accuracy.

\subsection{MIMO detection}
MIMO detection is a classical problem in simple hypothesis testing \cite{verdu1998multiuser}. The maximum
likelihood (ML) detector involves an exhaustive search and is 
the optimal detector in the sense of minimum joint probability of
error for detecting all the symbols simultaneously.  Unfortunately, it has an exponential runtime complexity which makes it impractical in large real time systems.

In order and overcome the computational cost of the maximum likelihood 
decoder there is considerable interest in implementation of
suboptimal detection algorithms which provide a better and more flexible accuracy vs complexity tradeoff.  
In the high accuracy regime, sphere decoding algorithms \cite{agrell2002closest}, \cite{guo2006algorithm}, \cite{suh2017reduced} were purposed, based on lattice search, and offering better computational complexity with a rather low accuracy performance degradation relatively to the full search. In the other regime, the most common suboptimal
detectors are the linear receivers, i.e., the matched filter (MF),
the decorrelator or zero forcing (ZF) detector and the minimum
mean squared error (MMSE) detector. More advanced detectors are
based on decision feedback equalization (DFE), approximate message passing (AMP) \cite{jeon2015optimality} and semidefinite relaxation (SDR) \cite{luo2010semidefinite,jalden2008diversity}. Currently, both AMP and SDR provide near optimal accuracy under many practical scenarios. AMP is simple and cheap to implement in practice, but is an iterative method that may diverge in challenging settings. SDR is more robust and has polynomial complexity, but is limited in the settings it addresses and is much slower in practice.

\subsection{Background on Machine Learning}

Machine learning is the ability to solve statistical problems using examples of inputs and their desired outputs. Unlike classical hypothesis testing, it is typically used when the underlying distributions are unknown and are characterized via sample examples. It has a long history but was previously limited to simple and small problems. Fast forwarding to recent years, the field witnessed the deep revolution. The ``deep'' adjective is associated with the use of complicated and expressive classes of algorithms, also known as architectures. These are typically neural networks with many non-linear operations and layers. Deep architectures are more expressive than shallow ones and can theoretically solve much harder and larger problems \cite{lecun2015deep}, but were previously considered impossible to optimize. With the advances in big data, optimization algorithms and stronger computing resources, such networks are currently state of the art in different problems from speech processing\cite{hinton2012deep,graves2014towards} and computer vision \cite{he2016deep,szegedy2015going} to online gaming \cite{silver2016mastering}. Typical solutions involve dozens and even hundreds of layers which are slowly optimized off-line over clusters of computers, to provide  accurate and cheap decision rules which can be applied in real-time.
In particular, one promising approach to designing deep architectures is by unfolding an existing iterative algorithm \cite{hershey2014deep}. Each iteration is considered a layer and the algorithm is called a network. The learning begins with the existing algorithm as an initial starting point and uses optimization methods to improve the algorithm. For example, this strategy has been shown successful in the context of sparse reconstruction \cite{gregor2010learning,borgerding2016onsager}. Leading algorithms as Iterative Shrinkage and Thresholding and a sparse version of AMP have both been improved by unfolding their iterations into a network and learning their optimal parameters. 

Following this revolution, there is a growing body of works on deep learning methods for communication systems. Exciting contributions in the context of error correcting codes include \cite{nachmani2016learning,nachmani2017rnn,nachmani2018deep,o2017introduction,gruber2017deep}. 
In \cite{farsad2017detection} a machine learning approach is considered in order to decode over molecular communication systems where chemical signals are used for transfer of information. In these systems an accurate model of the channel is impossible to find. This approach of decoding without CSI (channel state information) is further developed in \cite{farsad2018neural}. Machine learning for  channel estimation is considered in \cite{ye2017power,o2017learning}.
End-to-end detection over continuous signals is addressed in \cite{dorner2018deep}. And in \cite{o2017deep} deep neural networks are used for the task of MIMO detection using an end-to-end approach where learning is deployed both in the transmitter in order to encode the transmitted signal and in the receiver where unsupervised deep learning is deployed using an autoencoder. Parts of our work on MIMO detection using deep learning have already appeared in \cite{samuel2017deep}, see also \cite{wang2017deep}.
Similar ideas were discussed in \cite{diskin2017deep} in the context of robust regression. 

\subsection{Main contributions}
The main contribution of this paper is the introduction of two deep learning networks for MIMO detection. We show that, under a wide range of scenarios including different channels models and various digital constellations, our networks achieve near optimal detection performance with low computational complexity.

Another important result we show is their ability to easily provide soft outputs as required by modern communication systems. We show that for different constellations the soft output of our networks achieve accuracy comparable to that of the M-Best sphere decoder with low computational complexity.  

In a more general learning perspective, an important contribution is DetNet's ability to perform on multiple models with a single training. Recently, there were  works on learning to invert linear channels and reconstruct signals \cite{gregor2010learning,borgerding2016onsager,Mousavi2017Learning}. To the best of our knowledge, these were developed and trained to address a single fixed channel. In contrast, DetNet is designed for handling multiple channels simultaneously with a single training phase.

The paper is organized in the following order:

In section \RNum{2} we present the MIMO detection problem and how it is formulated as a learning problem including the use of one-hot representations.
In section \RNum{3} we present two types of neural network based detectors, FullyCon and DetNet.
In section \RNum{4} we consider soft decisions.
In section \RNum{5} we compare the accuracy and the runtime of the purposed learning based detectors against traditional detection methods both in the hard decision and the soft decision cases. Finally, section \RNum{6} provides concluding remarks.

\subsection{Notation}
In this paper, we define the normal distribution where $\mu$ is the mean and $\sigma^{2}$ is the variance as $\mathcal{N}\left(\mu,\sigma^{2}\right)$.  The uniform distribution with the minimum value $a$ and the maximum value $b$ will be $\mathcal{U}\left(a,b\right)$ .
Boldface uppercase letters denote matrices. Boldface lowercase letters denote vectors. The superscript $\left(\cdot\right)^{T}$ denotes the transpose. The i'th element of the vector $\x$ will be denoted as $\x_{i}$.
Unless stated otherwise, the term independent and identically distributed  (i.i.d.) Gaussian matrix, refers to a matrix where each of its elements is i.i.d. sampled from the normal distribution $\mathcal{N}\left(0,1\right)$.
The rectified linear unit defined as $\rho(x) = \max\{0,x\}$.
When considering a complex matrix or vector the real and imaginary parts of it are defined as $\Re(\cdot)$ and $\Im(\cdot)$ respectively. 
An $\alpha$-Toeplitz $\M$ matrix will be defined as a matrix such that $\M^{T}\M$ is a square matrix where the value of each element on the i'th diagonal is $\alpha^{i-1}$.

\section{Problem formulation}
\subsection{MIMO detection}
We consider the standard linear MIMO model:
\begin{equation}\label{linearmodel1}
 \bar{\y} = \bar{\H}\bar{\x} + \bar{\w},
\end{equation}
where $\mathbf{\bar{y}} \in \mathbb{C}^{N}$ is the received vector, $\mathbf{\bar{H}}\in \mathbb{C}^{N \times K}$ is the channel 
matrix, $\mathbf{\bar{x}} \in \mathbb{{\bar{S}}}^{K}$ is an unknown vector of independent and equal probability symbols from some finite constellation $\mathbb{\bar{S}}$ (e.g. PSK or QAM),
$\mathbf{\bar{w}}$ is a noise vector of size $N$ with independent, zero mean Gaussian variables of variance $\sigma^2$.

Our detectors do not assume knowledge of the noise variance $\sigma^2$.
Hypothesis testing theory guarantees that it is unnecessary for optimal detection. Indeed, the ML rule does not depend on it. This is contrast to the MMSE and AMP decoders that exploit this parameter and are therefore less robust in cases where the noise variance is not known exactly.

\subsection{Reparameterization}
A main challenge in MIMO detection is the use of complex valued signals and various digital constellations $\mathbb{\bar{S}}$ which are less common in machine learning. In order to use standard tools and provide a unified framework, we re-parameterize the problem using real valued vectors and one-hot mappings as described below. 

First, throughout this work, we avoid handling complex valued variables,  and use the following convention:
\begin{equation}\label{linearmodel2}
 \y = \H\x + \w,
\end{equation}
where
\begin{eqnarray}\label{complexdecouple}
\y
&=&
\begin{bmatrix}
   \Re(\bar{\y}) \\
   \Im(\bar{\y}) 
\end{bmatrix},
\w
=
\begin{bmatrix}
   \Re(\bar{\w}) \\
   \Im(\bar{\w}) 
\end{bmatrix},
\x
=
\begin{bmatrix}
   \Re(\bar{\x}) \\
   \Im(\bar{\x}) 
\end{bmatrix},
\nonumber
\\\H
&=&
\begin{bmatrix}
   \ \Re(\bar{\H}) \ \ -\Im(\bar{\H})\\
   \Im(\bar{\H}) \  \ \ \Re(\bar{\H})  
\end{bmatrix}
\end{eqnarray}
where $\mathbf{y} \in \mathbb{R}^{2N}$ is the received vector, $\mathbf{H}\in \mathbb{R}^{2N \times 2K}$ is the channel 
matrix and $\mathbf{x} \in \mathbb{S}^{2K}$ where $\mathbb{S} = \Re\{\mathbb{\bar{S}}\}$ (which is also equal to $\Im\{\mathbb{\bar{S}}\}$ in the complex valued constellations we tested)

A second convention  concerns the re-parameterization of the discrete constellations 
$\mathbb{{S}}=\{s_1,\cdots,s_{|\mathbb{{S}}|}\}$ using one-hot mapping. With each possible $s_i$ we associate a unit vector $\u_i\in \mathbb{R}^{|\mathbb{{S}}|}$. 
For example, the $4$ dimensional one-hot mapping of the real part of 16-QAM constellations is defined as
\begin{eqnarray}
  s_1=-3  &\leftrightarrow & \u_1=[1,0,0,0] \nonumber\\ 
   s_2=-1 &\leftrightarrow  & \u_2=[0,1,0,0] \nonumber\\
  s_3=1  &\leftrightarrow & \u_3=[0,0,1,0] \nonumber\\ 
  s_4=3  &\leftrightarrow & \u_4=[0,0,0,1] 
\end{eqnarray}
We denote this mapping via the function $s= f_{oh}(\u)$ so that $s_i=f_{oh}(\u_i)$ for $i=1,\cdots,|\mathbb{{S}}|$. More generally, for approximate inputs which are not unit vectors, the function is defined as
\begin{equation}
x=f_{oh}(\x_{oh})=\sum_{i=1}^{|\mathbb{S}|}s_i[\x_{oh}]_i
\end{equation}
The description above holds for a scalar symbol. The MIMO model involves a vector of $2K$ symbols which is handled by stacking the one-hot mapping of each of its elements. Altogether, a vector $\x_{oh} \in {\{0,1\}}^{|\mathbb{S}|\cdot 2K}$is mapped to  $\x\in \mathbb{{S}}^{2K}$.


\subsection{Learning to detect}
We end this section by formulating the MIMO detection problem as a machine learning task. 
The first step in machine learning is choosing a class of possible detectors, also known as an architecture. A network architecture is a function  $\hat{\x}_{oh}(\H,\y;\thet)$ parameterized by $\thet$ that detects the unknown $\x_{oh}$ given $\y$ and $\H$.
Learning is the problem of finding the $\thet$ within some feasible set that will lead to strong detectors $\hat{\x}_{oh}(\H,\y;\thet)$.
For this purpose, we fix a loss function $l\left(\x_{oh};\hat{\x}_{oh}\left(\H,\y;\thet\right)\right)$ that measures the distance between the true vectors and their estimates.
Then, we find the network's parameter  $\thet$ by minimizing the loss function over the MIMO model distribution:
\begin{eqnarray}\label{learning_min}
 \min_{\thet} \EE{ l\left(\x_{oh};\hat{\x}_{oh}(\H,\y;\thet)\right)},
\end{eqnarray}
where the expectation is with respect to all the random variables in (\ref{linearmodel2}), i.e., $\x$,  $\w$, and $\H$. Learning to detect is defined as finding the best parameters $\thet$ of the networks' architecture that minimize the expected loss $l\left(\cdot;\cdot\right)$ over the distribution in (\ref{linearmodel2}). 

We always assume perfect channel state information (CSI) which means that the channel $\H$ is exactly known during detection time. However, we differentiate between two possible cases:
\begin{itemize}
 \item Fixed Channel (FC): In the FC scenario, $\H$ is deterministic and constant (or a realization of a degenerate distribution which only takes a single value). This means that during the training phase we know over which channel the detector will detect.   
\item Varying Channel (VC): In the VC scenario, we assume $\H$ random with a known continuous distribution. It is still completely known but changes in each realization, and a single detection algorithm must be designed for all its possible realizations. When detecting, the channel is randomly chosen, and the network must be able to generalize over the entire distribution of possible channels.
\end{itemize}

Altogether, our goal is to detect $\x$, using a neural network that receives $\y$ and $\H$ as inputs and provides an estimate $\hat{\x}$.
In the next section, we will introduce two competing architectures that tradeoff accuracy and complexity.

\section{Deep MIMO detectors}
\subsection{FullyCon}
\begin{figure}[t]
  \center{\includegraphics[width=8.9cm]{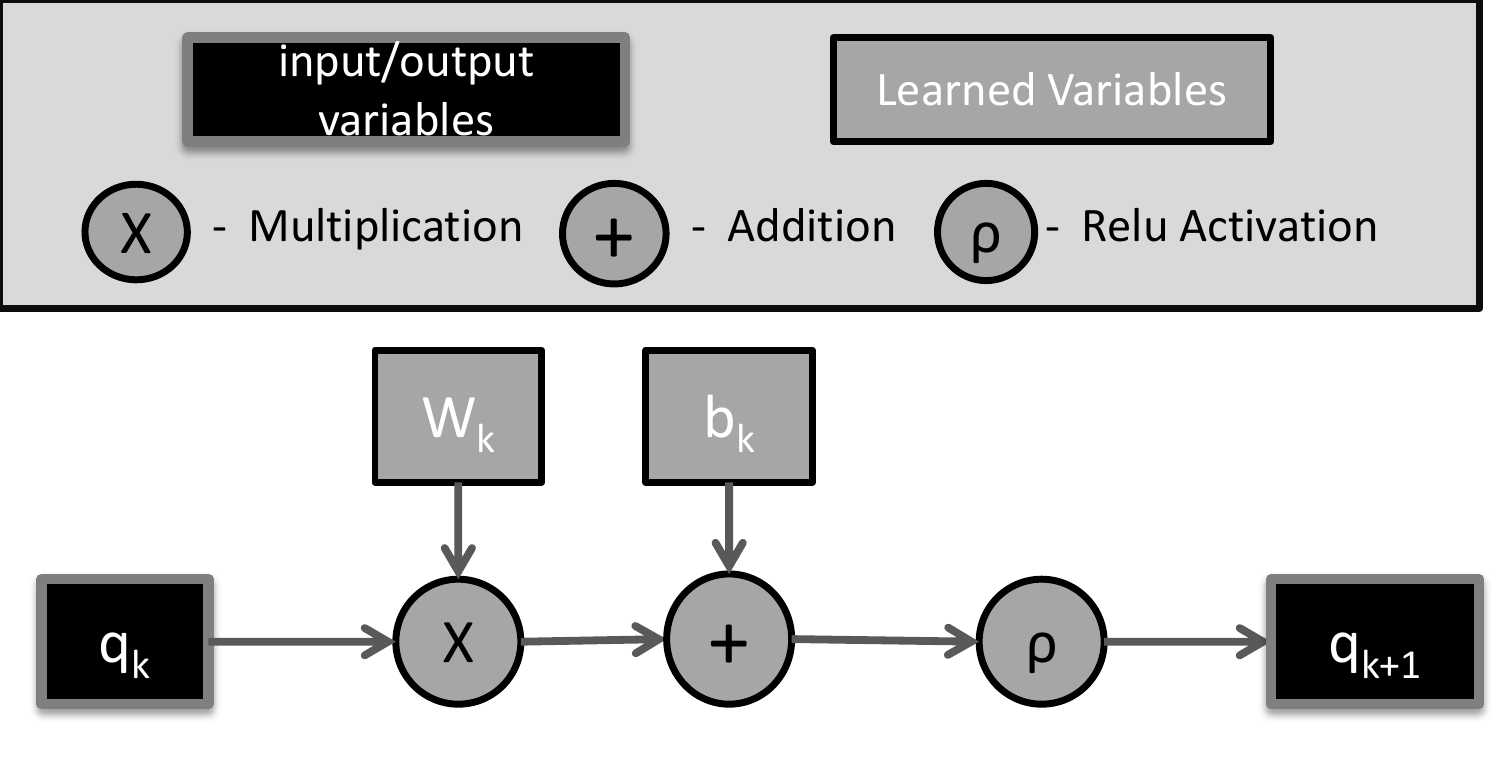}}
\caption{A flowchart representing a single layer of the fully connected network.}
\label{fig:FC_arch}%
\end{figure}

The fully connected multi-layer network is a well known architecture which is considered to be the basic deep neural network architecture, and from now on will be named simply as 'FullyCon'. It is composed of $L$ layers, where the output of each layer is the input of the next layer. Each layer can be described by the following equations:
\begin{eqnarray}\label{FC_Architecture}
\q_{1}&=&\y\nonumber \\ 
 \q_{k+1} &=& \rho\left(\W_{k}\q_{k}+\b_{k}\right) \nonumber\\
 \hat\x_{oh}&=&\W_{L}\q_L+\b_{L}\nonumber\\
 \hat\x&=&\f_{oh}(\hat\x_{oh})
\end{eqnarray}
An illustration of a single layer of FullyCon can be seen in Fig \ref{fig:FC_arch}.
The parameters of the network that are optimized during the learning phase are:
\begin{eqnarray}
 \thet=\left\{\W_{k},\b_{k}\right\}_{k=1}^L.
\end{eqnarray}
The loss function used is a simple $l_{2}$ distance between the estimated signal and the true signal:
\begin{eqnarray}\label{lossFC}
  l\left(\x_{oh};\hat{\x}_{oh}\left(\H,\y;\thet\right)\right) = 
\|\x_{oh}-\hat{\x}_{oh}\|^2
\end{eqnarray}

FullyCon is simple and general purpose. It has a relatively small number of parameters to optimize. It only uses the input $\y$, and does not exploit the channel $\H$ within (\ref{FC_Architecture}). The dependence on the channel is indirect via the expectation in (\ref{learning_min}) which depends on $\H$ and leads to parameters that depend on its moments. The result is a simple and straight forward structure which is ideal for detection over the FC model. As will be detailed in the simulations section, it manages to achieve almost optimal accuracy with low complexity. On the other hand, our experiences with FullyCon for the VC model led to disappointing results. It was not expressive enough to capture the dependencies of changing channels. We also tried to add the channel matrix $\H$ as an input, and this attempt failed too. In the next subsection, we propose a more expressive architecture specifically designed for addressing this challenge.

\subsection {DetNet}
\begin{figure*}[h]
  \center{\includegraphics[width=\textwidth,height=5.5cm]{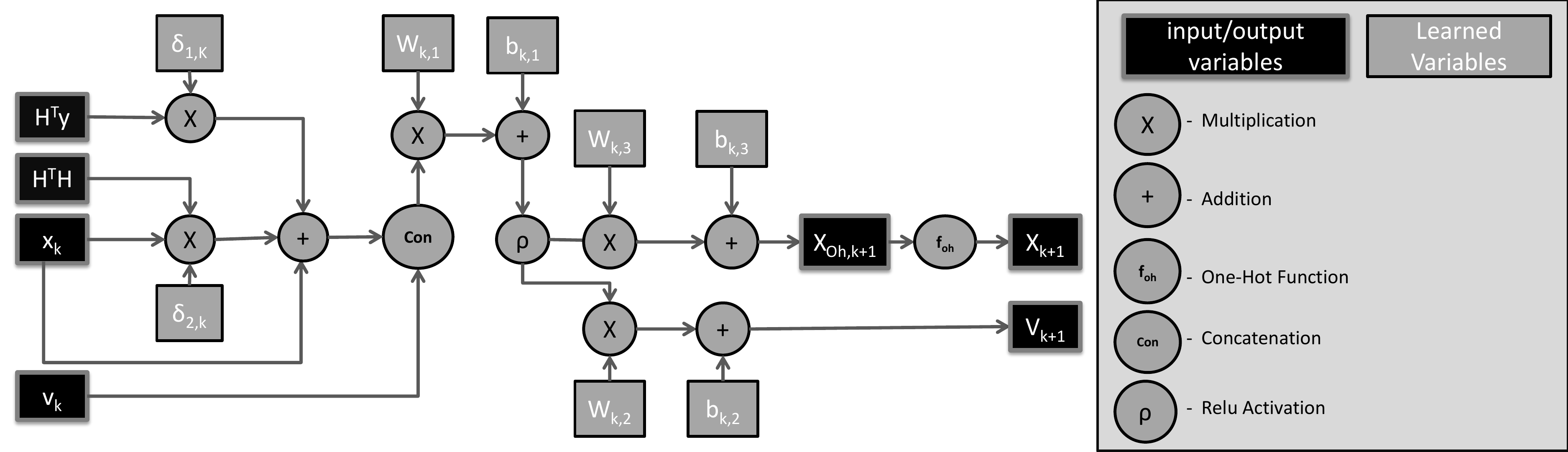}}
\caption{A flowchart representing a single layer of DetNet. The network is composed out of $L$ layers as such where each layers' output is the ext layers' input}
\label{fig:DetNet_arch}%
\end{figure*}

In this section we present an architecture designed specifically for MIMO detection that will be named from now on 'DetNet' (abbreviation of 'detection network'). The derivation begins by noting that an efficient MIMO detector should not work with $\y$ directly, but use the compressed sufficient statistic:
\begin{eqnarray}
 \H^T\y=\H^T\H\x+\H^T\w.
\end{eqnarray}
This hints that two main ingredients in the architecture should be $\H^T\y$ and $\H^T\H\x$.  Second, our construction is based on mimicking a projected gradient descent like solution for the maximum likelihood optimization. Such an algorithm would lead to iterations of the form
\begin{eqnarray}
 \hat\x_{k+1}&=&\Pi\left[\hat\x_k-\delta_k\left.\frac{\partial \|\y-\H\x\|^2}{\partial \x}\right|_{\x=\hat\x_k}\right]\nonumber\\
 &=&\Pi\left[\hat\x_k-\delta_k\H^T\y+\delta_k\H^T\H\x_k\right],
\end{eqnarray}
 where $\hat\x_k$ is the estimate in the $k$'th iteration, $\Pi[\cdot]$ is a nonlinear projection operator, and $\delta_k$ is a step size. Intuitively, each iteration is a linear combination of the $\x_k$, $\H^T\y$, and $\H^T\H\x_k$ followed by a non-linear projection. We enrich these iterations by lifting the input to a higher dimension in each iteration and applying standard non-linearities which are common in deep neural networks. In order to further improve the performance we treat the gradient step sizes $\delta_K$ at each step as a learned parameter and optimize them during the training phase.
 This yields the following architecture:
\begin{eqnarray}\label{DetNetArchitecture}
\nonumber \\
\q_{k}&=&\hat\x_{k-1}-\delta_{1k}\H^T\y+\delta_{2k}\H^T\H\x_{k-1}\nonumber \\ \z_{k} &=& \rho\left(\W_{1k} \left[
 \begin{array}{c}
   \q_{k} \\   \v_{k-1} 
 \end{array} \right] 
 +\b_{1k}\right) \nonumber\\
 \hat\x_{oh,k}&=&\W_{2k}\z_k+\b_{2k}\nonumber\\
 \hat\x_{k}&=&\f_{oh}(\hat\x_{oh,k})\nonumber\\
 \hat\v_{k}&=&\W_{3k}\z_k+\b_{3k} \nonumber\\
 \hat\x_0 &=& \0\nonumber\\
 \hat\v_0 &=& \0,
\end{eqnarray}
with the trainable parameters
\begin{eqnarray}
 \thet=\left\{\W_{1k},\b_{1k},\W_{2k},\b_{2k},\W_{3k},\b_{1k},\delta_{1k},\delta_{2k}\right\}_{k=1}^L.
\end{eqnarray}
To enjoy the lifting and non-linearities, the parameters $\W_{1k}$ are defined as tall and skinny matrices. The final estimate is defined as $\hat\x_L$. For convenience, the structure of each DetNet layer is illustrated in Fig. \ref{fig:DetNet_arch}.

Training  deep networks is a difficult task due to vanishing gradients, saturation of the activation functions, sensitivity to initialization and more \cite{glorot2010understanding}. To address these challenges and following the notion of auxiliary classifiers feature in GoogLeNet \cite{szegedy2015going}, we adopted a loss function that takes into account the outputs of all of the layers:
\begin{eqnarray}
  l\left(\x_{oh};\hat{\x}_{oh}\left(\H,\y;\thet\right)\right) = \sum_{l=1}^{\L} \log(l)
{\|\x_{oh}-\hat{\x}_{oh,l}\|^2}.
\end{eqnarray}

In our final implementation, in order to further enhance the performance of DetNet, we added a residual feature from ResNet \cite{he2016deep} where the output of each layer is a weighted average with the output of the previous layer. 

\section{Soft decision output}
In this section, we consider a more general setting in which the MIMO detector needs to provide soft outputs. High end communication systems typically resort to iterative decoding where the MIMO detector and the error correcting decoder iteratively exchange information on the unknowns until convergence. For this purpose, the MIMO detector must replace its hard estimates with soft posterior distributions Prob$(x_j=s_i|\y)$ for each unknown $j=1,\cdots,2K$ and each possible symbol $i=1,\cdots,|\mathbb{S}|$. More precisely, it also needs to allow additional soft inputs but we leave this for future work. 


Computation of the posteriors is straight forward based on Bayes law, but its complexity is exponential in the size of the signal and constellation. Similarly to the maximum likelihood algorithm in the hard decision case, this computation yields optimal accuracy yet is intractable.  Thus, the goal in this section is to design networks that output approximate the posteriors. On first glance, this seems difficult to learn as we have no training set of posteriors and cannot define a loss function. Remarkably, this is not a problem and the probabilities of arbitrary constellations can be easily recovered using the standard $l_2$ loss function with respect to the one-hot representation $x_{oh}$. Indeed, consider a scalar $x$ and a single $s\in \mathbb{S}$ associated with its one-hot bit $x_{oh}$ then it is well known that 
\begin{eqnarray}
{\rm{arg}}\min_{\hat{x}_{oh}}E[{||x_{oh}-\hat{x}_{oh}||}^2|\y] 
&=& E[x_{oh}|\y]\\&=&\rm{\underset{s \in \mathbb{S} }{Prob}(x_{oh,s}=1|\y)}\nonumber\\
&=&\rm{\underset{s \in \mathbb{S} }{Prob}}(x=s|\y)\nonumber
\end{eqnarray}
Thus, assuming that our network is sufficiently expressive and globally optimized, the one-hot output $\hat{\x_{oh}}$ will provide the exact posterior probabilities.

\section{Numerical Results}
In this section, we provide numerical results on the accuracy and complexity of the proposed networks in comparison to competing methods. 

In the FC case, the results are over the 0.55-Toeplitz channel.

In the VC case and when testing the soft output performance, the results presented are over random channels, where each element is sampled i.i.d. from the normal distribution $\mathcal{N}\left(0,1\right)$.
\subsection{Implementation details}
We train both networks using a variant of the stochastic gradient descent method \cite{rumelhart1988learning},\cite{bottou2010large} for optimizing deep networks, named Adam Optimizer \cite{kingma2014adam}.
All networks were implemented using the Python based TensorFlow library \cite{abadi2016tensorflow}.
 
To give a rough idea of the computation needed during the learning phase, optimizing the detectors in our numerical results in both architectures took around 3 days on a standard Intel i7-6700 processor.  Each sample was independently generated from (\ref{linearmodel2}) according to the statistics of $\x$, $\H$ (either in the FC or VC model) and $\w$. During training, the noise variance was randomly generated so that the SNR will be uniformly distributed on $\mathcal{U}\left({\rm{SNR}}_{\min},{\rm{SNR}}_{\max}\right)$.
\subsection{Competing algorithms}
When presenting our network performance we shall use the following naming conventions:
\begin{description}
 \item [FullyCon:] The basic fully-connected deep architecture.
 \item [DetNet:] The DetNet deep architecture.
 \end{description}
 
In the hard decision scenarios, we tested our deep networks against the following detection algorithms:
\begin{description}
\item  [ZF:] This is the classical decorrelator, also known as least squares or zero forcing (ZF) detector \cite{verdu1998multiuser}.
 \item  [AMP:] Approximate message passing algorithm from \cite{jeon2015optimality}.
 \item  [SDR:] A decoder based on semidefinite relaxation implemented using an efficient interior point solver \cite{luo2010semidefinite,jalden2008diversity}. For the 8-PSK constellation we implemented the SDR variation suggested in \cite{ma2004semidefinite}.
  \item [SD:] An implementation of the sphere decoding algorithm as presented in \cite{ghasemmehdi2011faster}.
 \end{description}

In the soft output case, we tested our networks against the M-Best sphere decoding algorithm as presented in \cite{guo2006algorithm} (originally named K-Best, but changed here to avoid confusion with $K$ the transmitted signal size):
\begin{description}
 \item  [M-Best SD M=5:] The M-Best sphere decoding algorithm, where the number of candidates we keep is 5.
 \item  [M-Best SD M=7:] Same as M-Best SD M=5 with 7 candidates.
 \end{description}

\subsection{Accuracy results}
\subsubsection{Fixed Channel (FC)}
\begin{figure}[t]
  \center{\includegraphics[width=8.5cm]{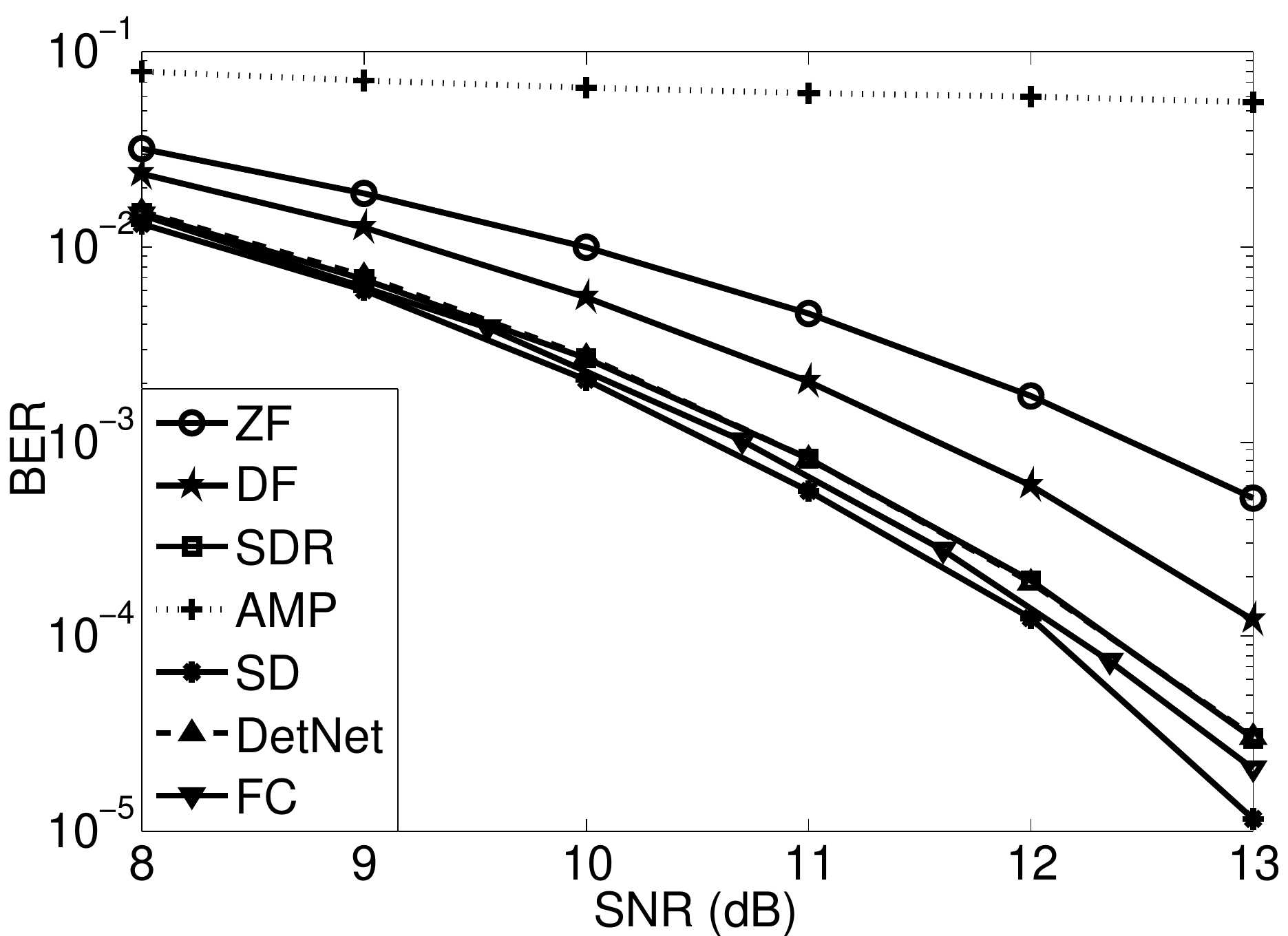}}
\caption{Comparison of the detection algorithms BER performance in the fixed channel channel case over a BPSK modulated signal.}
\label{fig:constmat055bpsk}
\end{figure}

In the case of the FC scenario, where we know during the learning phase over what realization of the channel we need to detect, the performance of both our network was comparable to most of the competitors except SD. Both  DetNet and FullyCon  managed to achieve accuracy results comparable to SDR and AMP.
This result emphasizes the notion that when learning to detect over simple scenarios as FC, a simple network is expressive enough. And since a simple network is easier to optimize and has lower complexity, it is preferable.
Fig. \ref{fig:constmat055bpsk} we present the accuracy rates over a range of SNR values in the FC model. This is a rather difficult setting and algorithms such as AMP did not succeed to converge.
\subsubsection{Varying channel}
\begin{figure}[t]
  \center{\includegraphics[width=8.5cm]{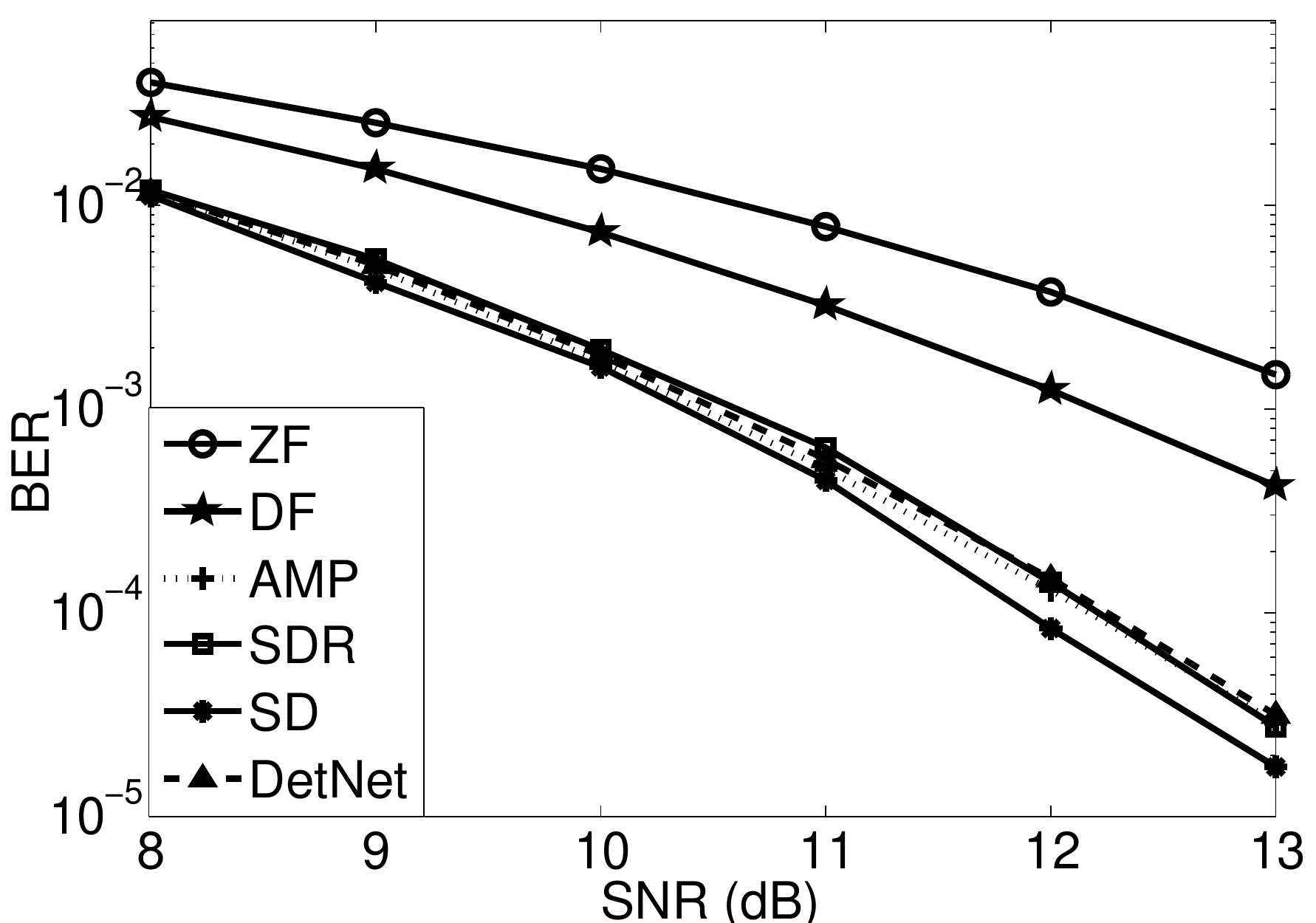}}
\caption{Comparison of the detection algorithms BER performance in the varying channel case over a BPSK modulated signal. All algorithms were tested channels of size 30x60.}
\label{fig:bpsk}
\end{figure}

\begin{figure}[t]
  \center{\includegraphics[width=8.5cm]{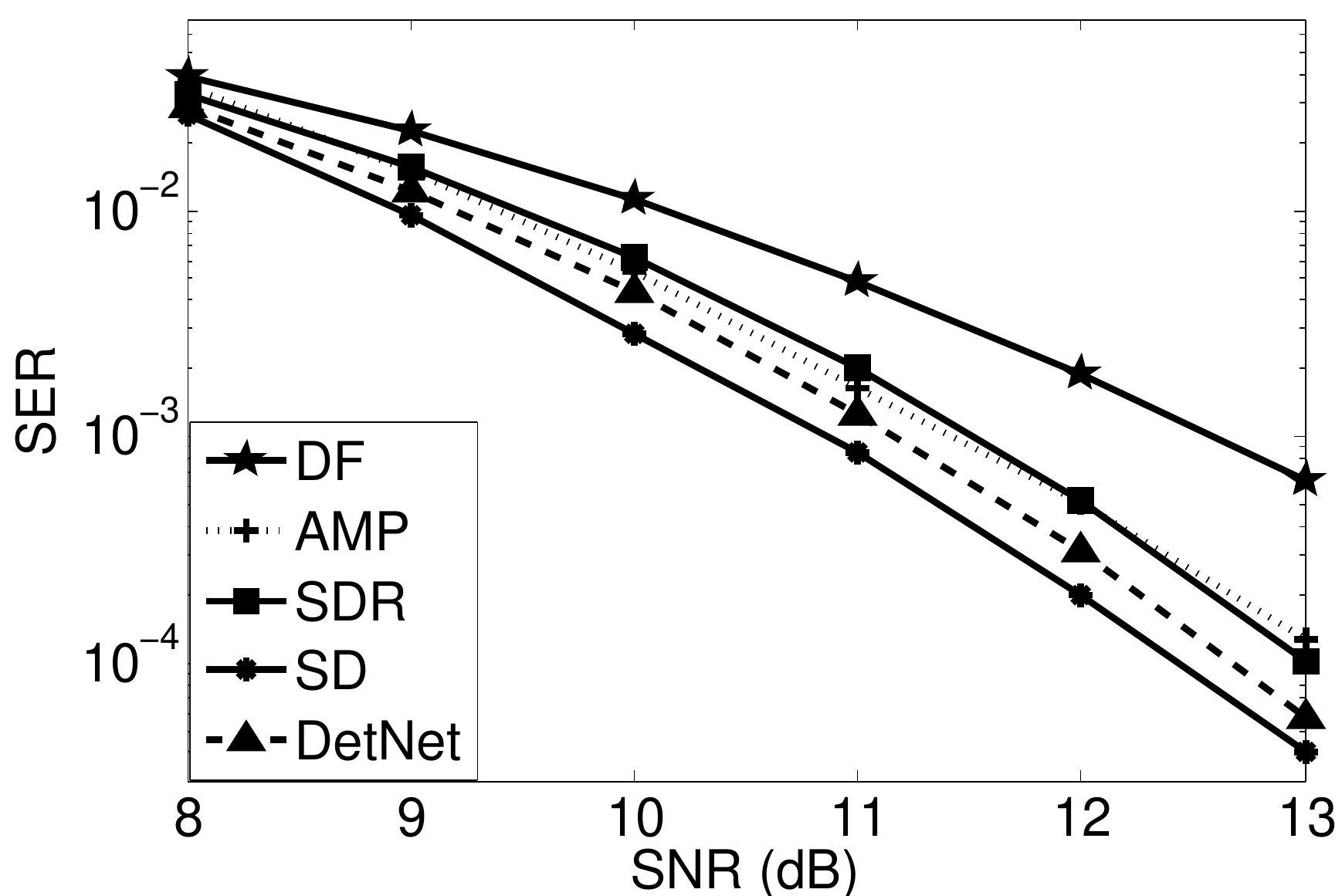}}
\caption{Comparison of the detection algorithms BER performance in the varying channel case over a QPSK modulated signal. All algorithms were tested on channels of size 20x30.}
\label{fig:qpsk}
\end{figure}

\begin{figure}[t]
  \center{\includegraphics[width=8.5cm]{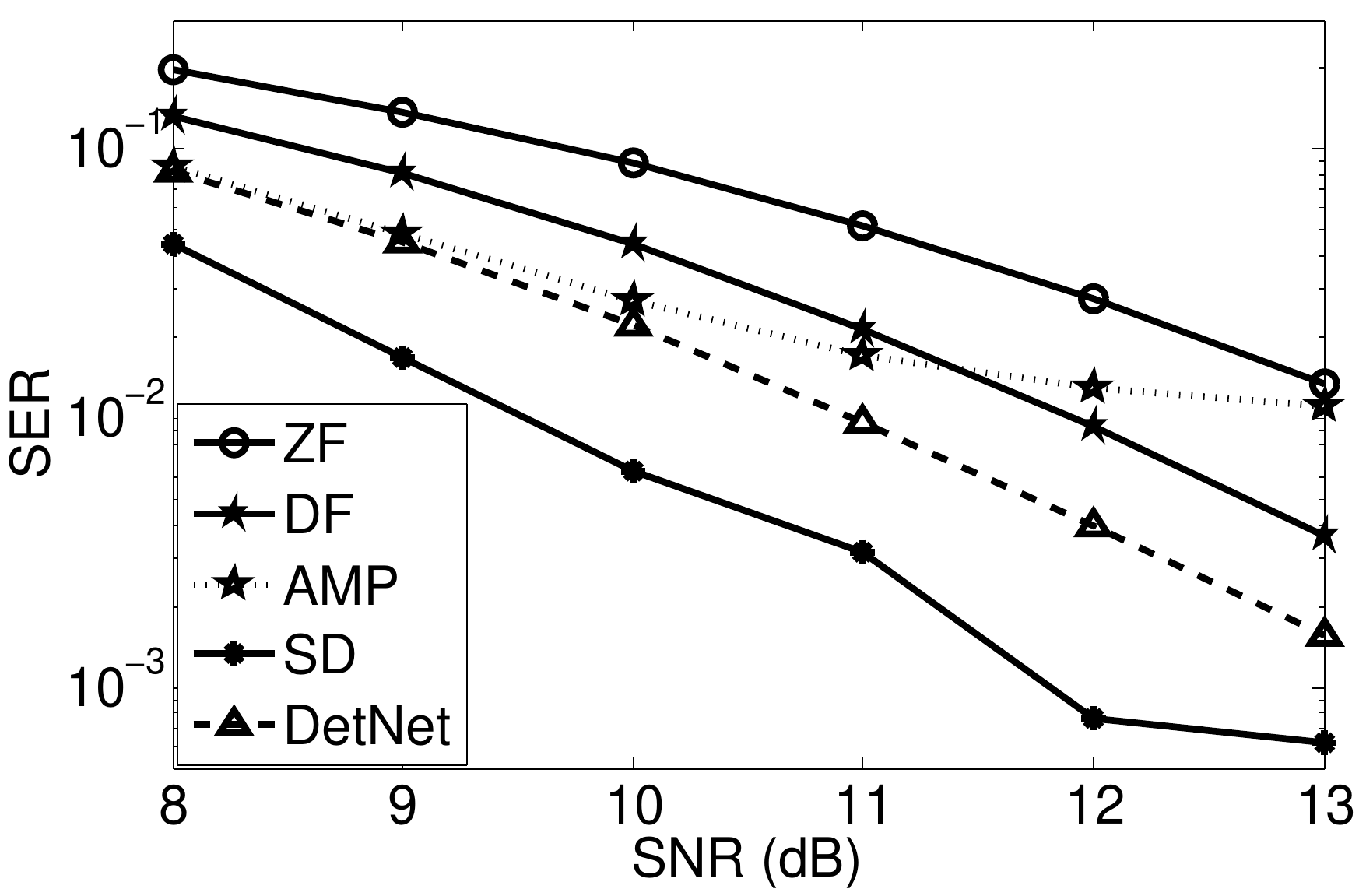}}
\caption{Comparison of the detection algorithms SER performance in the varying channel case over a 16-QAM modulated signal. All algorithms were tested on channels of size 15X25.}
\label{fig:16QAM}
\end{figure}

\begin{figure}[t]
  \center{\includegraphics[width=8.5cm]{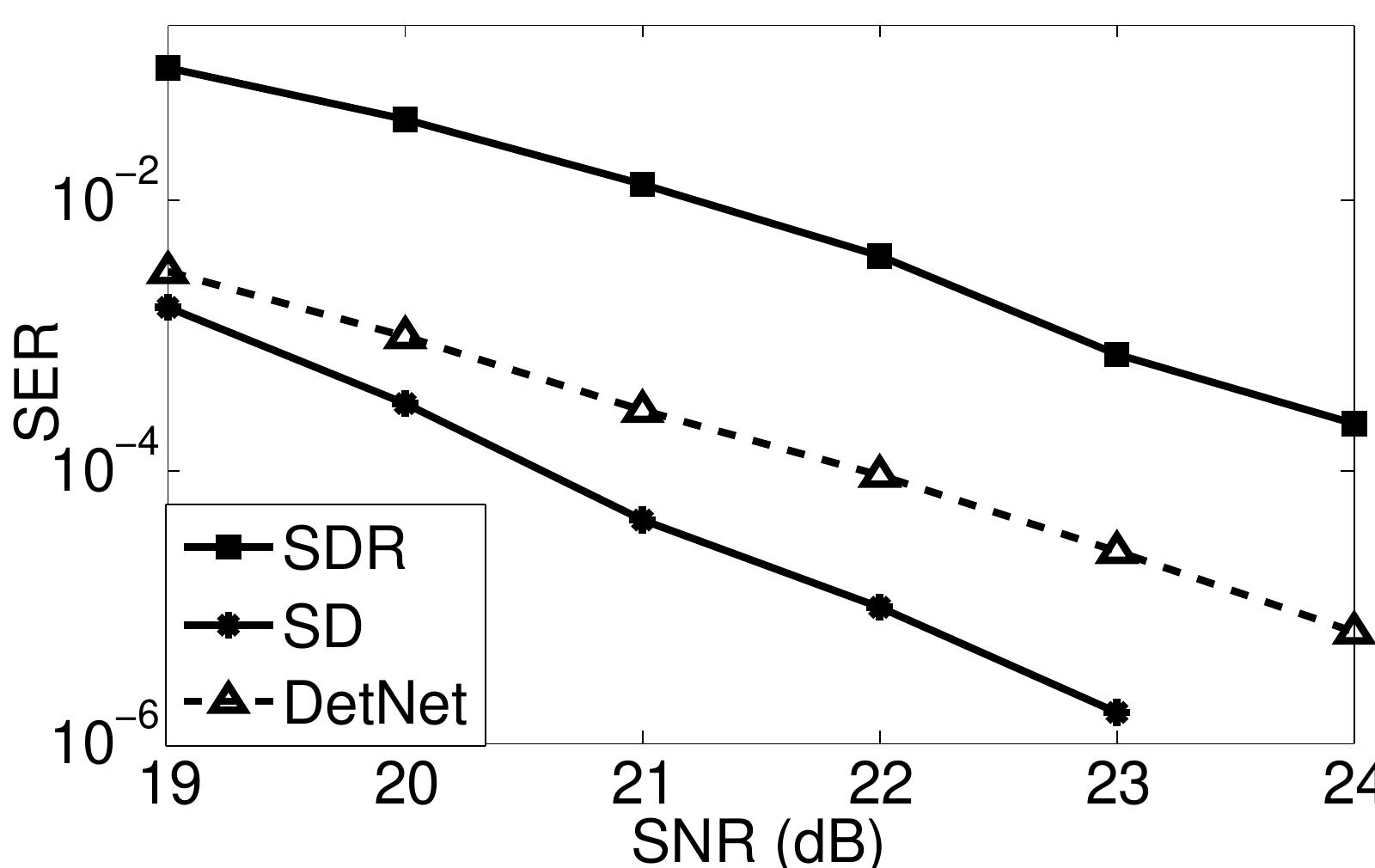}}
\caption{Comparison of the detection algorithms SER performance in the varying channel case over a 8-PSK modulated signal. All algorithms were tested on channels of size 15X25.}
\label{fig:8psk}
\end{figure}

In the  VC case, the accuracy results of FullyCon were poor and the network did not manage to learn how to detect properly. DetNet managed to achieve accuracy rates comparable to those of SDR and AMP, and almost comparable to those of SD, while being computationally cheaper (see next section regarding computational resources).
In Fig. \ref{fig:bpsk} we compare the accuracy results over a $30\times60$ real valued channel with BPSK signals and in Fig. \ref{fig:qpsk} we compare the accuracy of a $20\times30$ complex channel with  QPSK symbols. In both cases DetNet achieves accuracy rates comparable to SDR and AMP and near SD, and accuracy much better than ZF and DF.
Results over larger constellations are presented in Fig. \ref{fig:16QAM} and \ref{fig:8psk} where we compare the accuracy rates over complex channels of size $15\times25$ for the 16-QAM and 8-PSK constellations respectively.We can see that in those larger constellations DetNet performs better then AMP and SDR. For both constellations we can observe that DetNet reaches accuracy levels topped only by SD.

\subsubsection{Soft Outputs}
We also experimented with soft decoding. Implementing a full iterative decoding scheme is outside the scope of this paper, and we only provide initial results on the accuracy of our posterior estimates. For this purpose, we examined smaller models where the exact posteriors can be computed exactly and measured their statistical distance to our estimates.

\begin{figure}[t]
\begin{minipage}[b]{.8\linewidth}
  \center{\includegraphics[width=8.5cm]{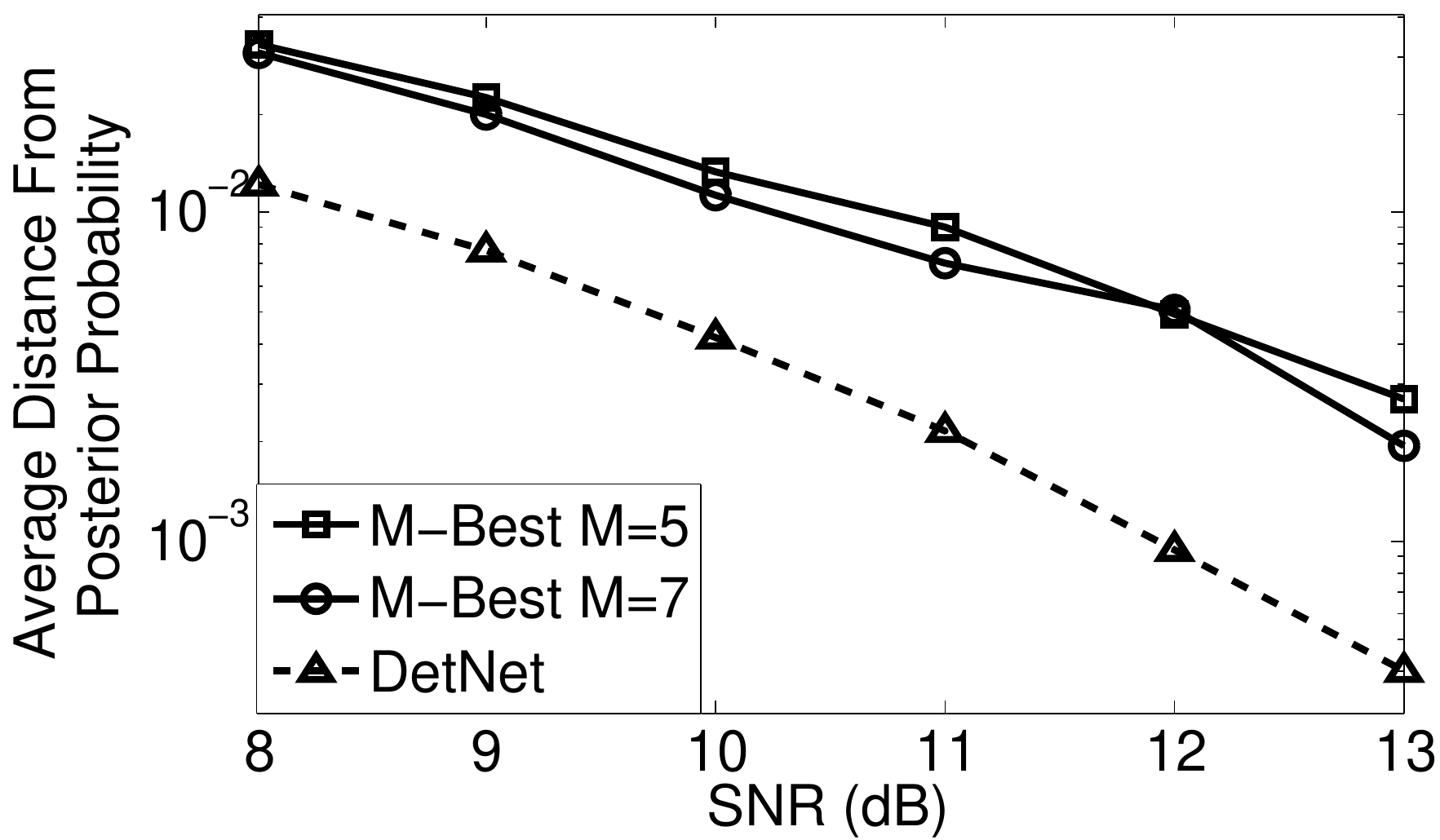}}
\end{minipage}
\caption{Comparison of the accuracy of the soft output relative to the posterior probability in the case of a BPSK signal over a $10\times20$ real valued channel. We present the results for 2 types of DetNet, one with 30 layers and the second one with 50 layers}
\label{fig:Soft_BPSK_10_20}
\end{figure}

\begin{figure}[t]
\begin{minipage}[b]{.8\linewidth}
  \center{\includegraphics[width=8.5cm]{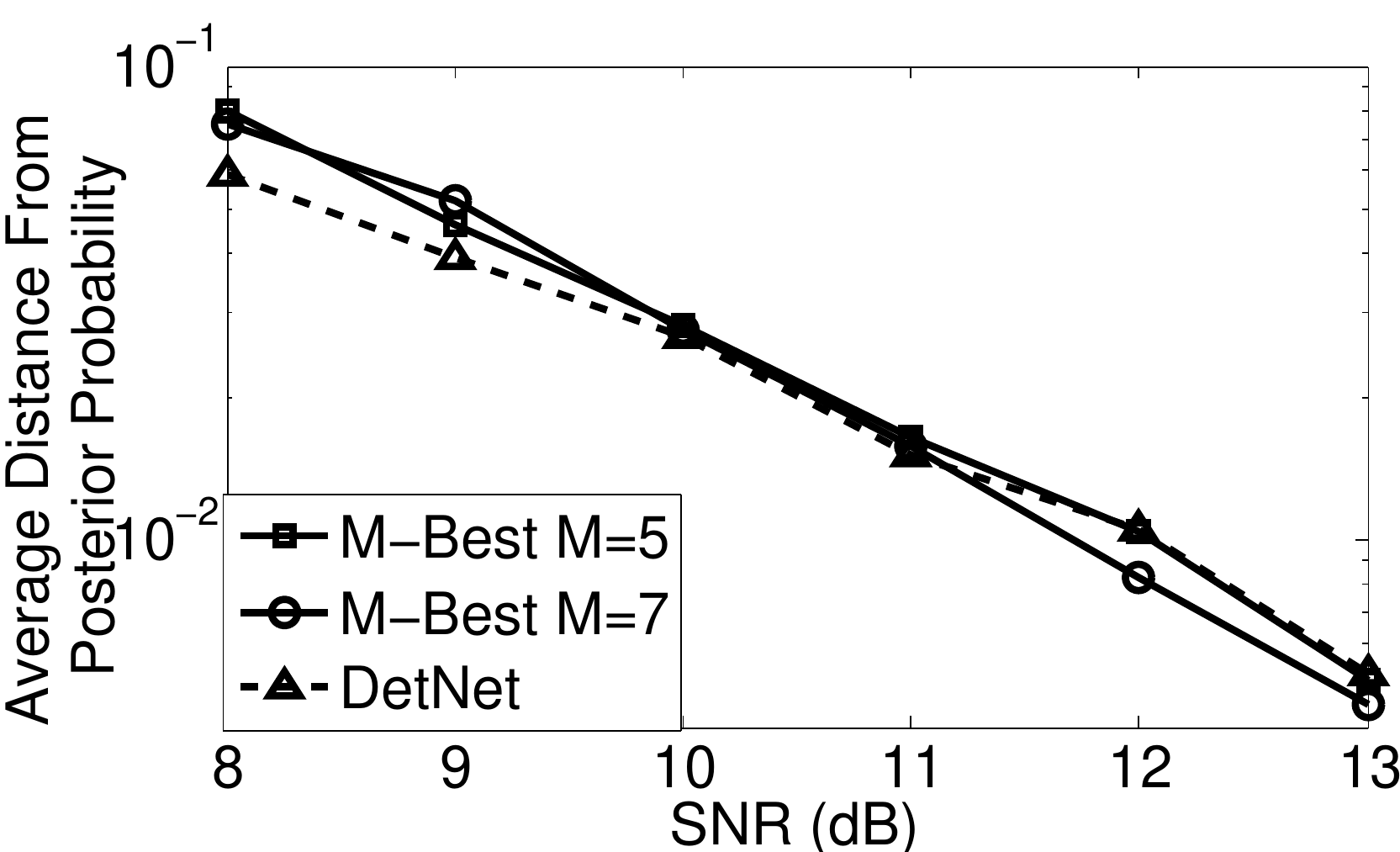}}
\end{minipage}
\caption{Comparison of the accuracy of the soft output relative to the posterior probability for a 16-QAM signal over an  $4\times8$ complex valued channel.}
\label{fig:Soft_16QAM_4_8}
\end{figure}

\begin{figure}[t]
\begin{minipage}[b]{.8\linewidth}
  \center{\includegraphics[width=8.5cm]{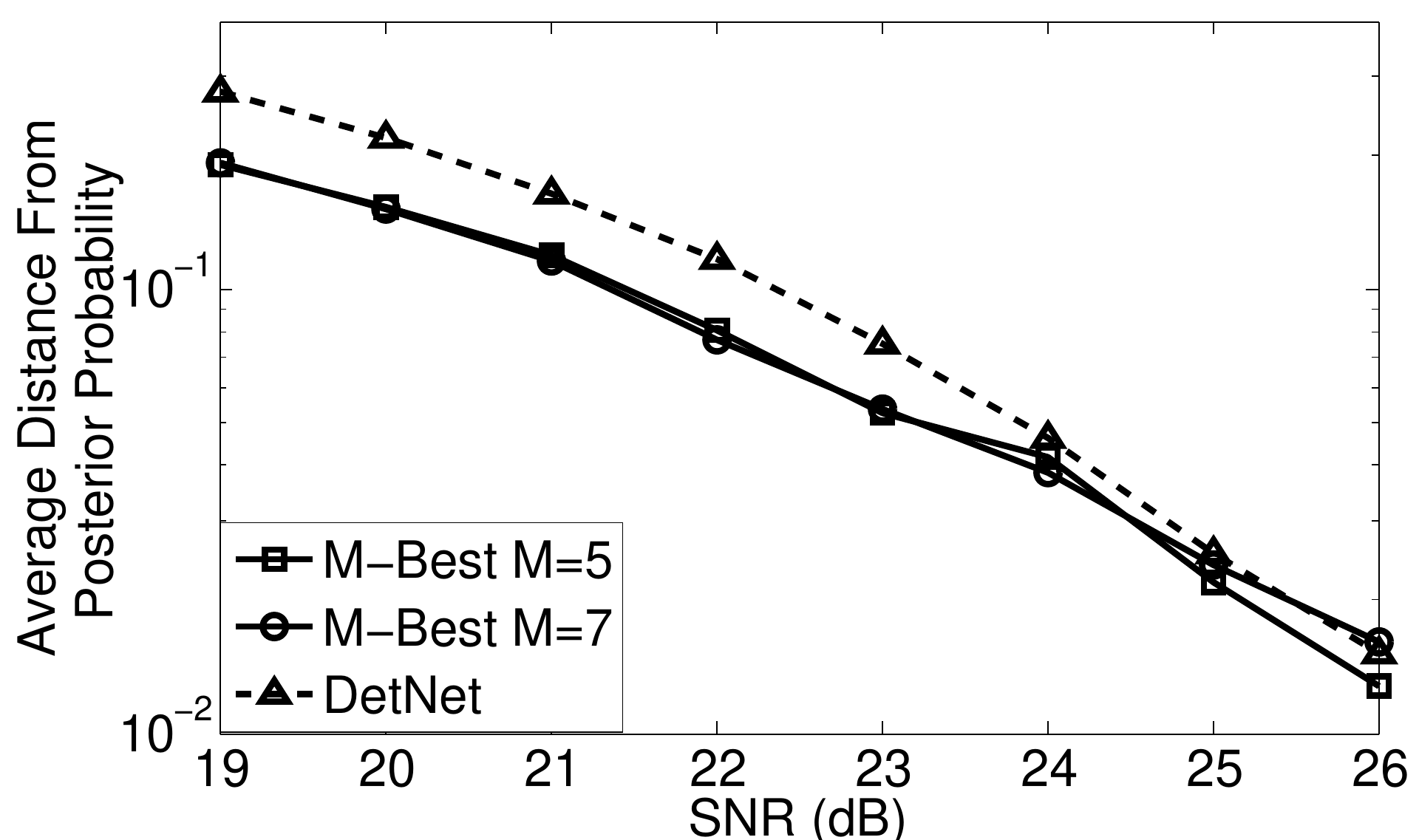}}
\end{minipage}
\caption{Comparison of the accuracy of the soft output relative to the posterior probability for a 8-PSK signal over an  $4\times8$ complex valued channel.}
\label{fig:Soft_8PSK_4_8}
\end{figure}

We shall define the following statistical distance function:

Given two probability distributions $P$ and $Q$ over the symbol set $\mathbb{{S}}$ (that is, the probability of each symbol to be the true symbol), the distance $\delta(P,Q)$ shall be:
\begin{eqnarray}
\delta(P,Q) = \sum_{s\in \mathbb{{S}}} |P(s)-Q(s)|
\end{eqnarray}
As reference, we compare our results to the M-Best detectors \cite{guo2006algorithm}.
In Fig. \ref{fig:Soft_BPSK_10_20} we present accuracy in the case of a BPSK signal over a 10x20 real channel. In this setting we reach accuracy levels better than those achieved by the M-Best algorithm. As seen in Fig. \ref{fig:Soft_BPSK_10_20} adding additional layers improves the accuracy of the soft output.
In Fig. \ref{fig:Soft_16QAM_4_8} we present the results over a 4x8 complex channel with 16-QAM constellation. We can see the performance of DetNet is comparable to the M-Best Sphere decoding algorithm. For completeness, in Fig. \ref{fig:Soft_8PSK_4_8} we added the 8-PSK constellation soft output where DetNet is comparable to the M-Best algorithms only in the high SNR region.

\subsection{Computational Resources}
\subsubsection{FullyCon and DetNet run time}
In order and estimate the computational complexity of the different detectors we compared their run time. Comparing complexity is non-trivial due to many complicating factors as implementation details and platforms.  To ensure fairness, all the algorithms were tested on the same machine via python 2.7 environment using the Numpy package. The networks were converted from TensorFlow objects to Numpy objects. We note that the run-time of SD depends on the SNR, and we therefore report a range of times. 

An important factor when considering the run time of the neural networks is the effect the batch size. Unlike classical detectors as SDR and SD, neural networks can detect over entire batches of data which speeds up the detection process. This is true also for the AMP algorithm, where computation can be made on an entire batch of signals at once. However, the improvement introduced by using batches is highly dependent on the platform used (CPU/GPU/FPGA etc).  Therefore, for completeness, we present the run time for several batch sizes including batch size equal to one.

In table \ref{table:FCruntume} the run times are presented for hard decision detection in a FC case. We can see that  FullyCon  is faster than all other detection algorithms, even without using batches. DetNet is slightly faster than traditional detection algorithms without using batches, yet when using batches, the run time improves significantly compared to other detection methods.

\begin{table}[h!]
\begin{center}
 \begin{tabular}{||c | c| c | c | c | c | c ||} 
 \hline
 Channel      &  Batch & FullyCon &  DetNet & SDR   &   AMP    & SD  \\ 
  size        &  size  &          &         &       &          &     \\ [0.5ex] 
 \hline\hline
 Top055       & 1      & 0.0004   &  0.0045 & 0.009 &  0.005   & 0.001   \\ 
 30x60        &        &          &         &       &          & -0.01   \\ 
 \hline
 Top055       & 10     & 6.6E-05  &  0.0007 & 0.009 &  0.001   & 0.001   \\ 
 30x60        &        &          &         &       &          & -0.01   \\ 
 \hline
 Top055       & 100    & 2.4E-05  & 1.6E-04 & 0.009 & 0.0003   & 0.001   \\ 
 30x60        &        &          &         &       &          & -0.01   \\ 
 \hline
 Top055       & 1000   & 1.6E-05  & 1.1E-04 & 0.009 & 0.0003   & 0.001   \\ 
 30x60        &        &          &         &       &          & -0.01   \\ 
\hline
\end{tabular}
\caption{Fixed Channel Runtime Comparison}
\label{table:FCruntume}
\end{center}
\end{table}

In table \ref{table:runtime_hard_detnet} we present the results for the VC setting. In the BPSK case the relative time difference between the different detection algorithms is similar to the FC case, with the exception of SD being relatively slower. In larger constellations (8-PSK/16-QAM) DetNet's relative advantage when comparing against AMP/SDR is smaller than in the BPSK case (and in the 16-QAM constellation AMP was slightly faster without using batches). The reason is that these accurate detection with these constellations requires larger networks. On the other hand, the relative performance vs SD improved.

\begin{table}[h!]
\begin{center}
 \begin{tabular}{||c | c | c | c | c | c ||} 
 \hline
 Constellation & Batch &  DetNet & SDR   &   AMP   &    SD    \\
 channel size  & size  &         &       &         &          \\ [0.5ex] 
 \hline\hline
 BPSK     & 1     & 0.0066  & 0.024 & 0.0093  & 0.008  \\ 
 30X60    &       &         &       &         &  -0.1  \\     
 \hline
 BPSK     &  10   & 0.0011  & 0.024 & 0.0016  & 0.008 \\ 
 30X60    &       &         &       &         &  -0.1  \\ 
 \hline
 BPSK     & 100   & 0.0005  & 0.024 & 0.00086 & 0.008  \\ 
 30X60    &       &         &       &         &  -0.1  \\ 
 \hline
 16-QAM   & 1     & 0.006   &   -   & 0.01    & 0.01  \\
 15X25    &       &         &       &         & -0.4 \\ 
 \hline
 16-QAM   & 10    & 0.0014  &   -   & 0.002   & 0.01  \\
  15X25   &       &         &       &         & -0.4 \\ 
 \hline
 16-QAM   & 100   & 0.0003  &   -   & 0.001   & 0.01  \\
  15X25   &       &         &       &         & -0.4 \\ 
 \hline
 8-PSK    & 1     & 0.019   & 0.021 &   -     & 0.004 \\
 15X25    &       &         &       &         & -0.06 \\
 \hline
  8-PSK   & 10    & 0.0029  & 0.021 &   -     & 0.004 \\
   15X25  &       &         &       &         & -0.06 \\
 \hline
  8-PSK   & 100   & 0.0005  & 0.021 &   -     & 0.004 \\
   15X25  &       &         &       &         & -0.06 \\
 \hline
\end{tabular}
\caption{Run Time Comparison in VC. DetNet  is compared with the SDR,AMP and Sphere Decoding algorithms}
\label{table:runtime_hard_detnet}
\end{center}
\end{table}

In table \ref{table:runtume_soft} we compare the run time of the detection algorithms in the soft-output case.
As we can see, in the BPSK case without using batches the performance of DetNet is comparable to the performance of the M-Best sphere decoders, and using batches improves the performance significantly. In the 16-QAM/8-PSK cases DetNet is slightly faster than the M-Best decoders even without using batches.

\begin{table}[h!]
\begin{center}
 \begin{tabular}{||c | c | c | c | c ||} 
 \hline
Constellation  & Batch & DetNet    & M-Best     & M-Best  \\
channel size   & size  &           & (M=5)      & (M=7)   \\ [0.5ex] 
\hline\hline
   BPSK  10X20 & 1     &  0.0075   &  0.006     & 0.008   \\
\hline
   BPSK  10X20 & 10    &  0.00092  &  0.006     & 0.008   \\
\hline
   BPSK  10X20 & 100   &  0.00029  &  0.006     & 0.008   \\
\hline
   16-QAM 4X8   & 1     &  0.006   &  0.008     & 0.01    \\
\hline
   16-QAM 4X8   & 10    &  0.0008  &  0.008     & 0.01    \\
\hline
   16-QAM 4X8   & 100   &  0.0001  &  0.008     & 0.01    \\
\hline
   8-PSK  4X8   & 1     &  0.02    &  0.05      & 0.07    \\
\hline
   8-PSK  4X8   & 10    &  0.003   &  0.05      & 0.07    \\
\hline
   8-PSK  4X8   & 100   &  0.0012  &  0.05      & 0.07    \\
\hline
\end{tabular}
\caption{Run Time Comparison of Soft Output in VC. The DetNet  is compared with the M-Best Sphere Decoding algorithm}
\label{table:runtume_soft}
\end{center}
\end{table}

\subsubsection{Accuracy-Complexity Trade-Off}
\begin{figure}[t]
\begin{minipage}[b]{.8\linewidth}
  \center{\includegraphics[width=8.5cm]{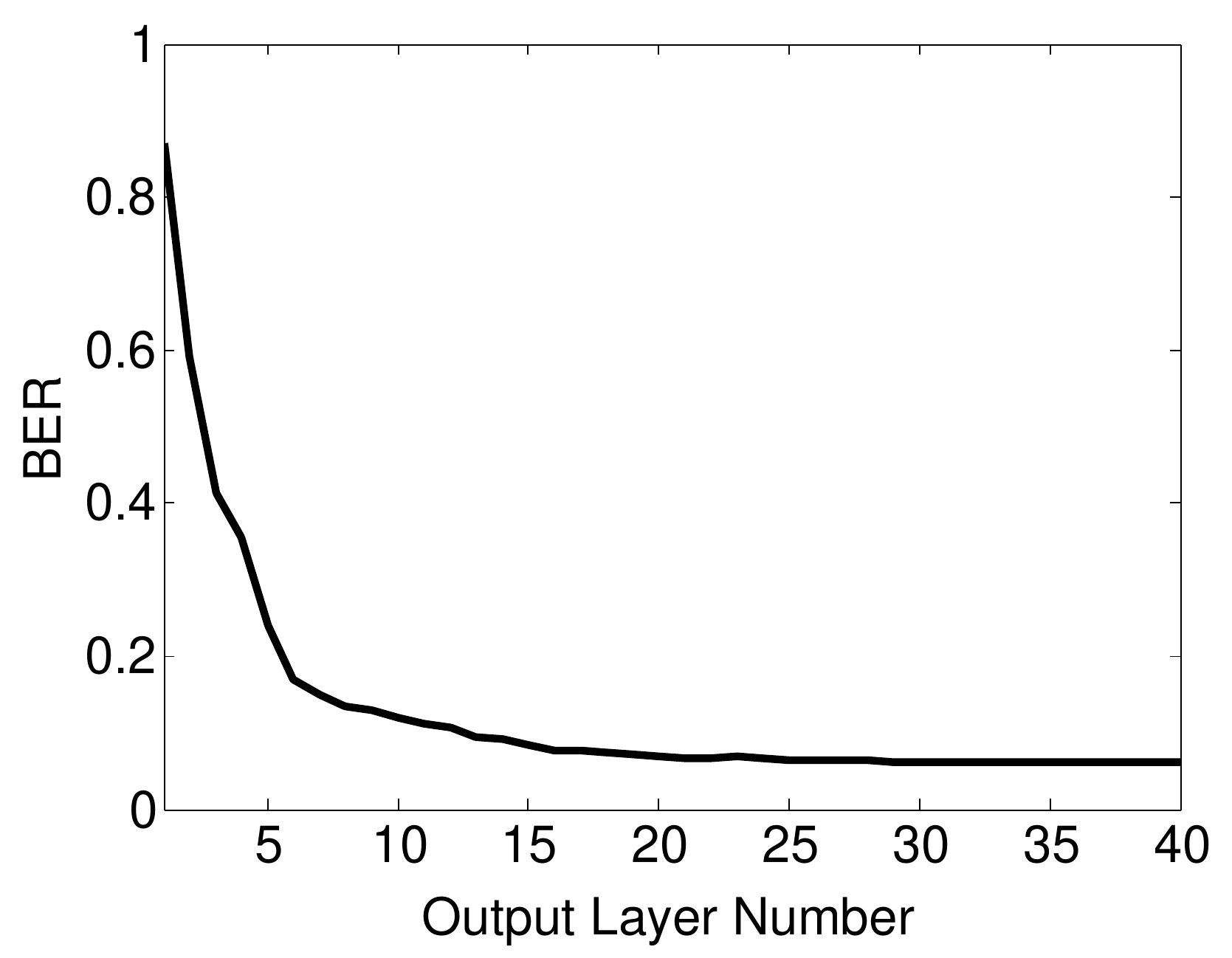}}
\end{minipage}
\caption{Comparison of the average BER as a function of the layer chosen to be the output layer. 
}
\label{fig:bers_layers}
\end{figure}

An interesting feature of DetNet is that  the complexity-accuracy trade-off can be decided during run-time.
Each of the network's layers outputs an estimated signal, and our loss optimizes all of them. We usually use the output of the last layer as the result since it is the most accurate, but it is possible to take the estimated output $\x_{i}$ of previous layers to allow faster detection. 
In Fig. \ref{fig:bers_layers} we present the accuracy as a function of the number of layers.

\section{Conclusion}
In this paper we investigated into the ability of deep neural networks to serve as MIMO detectors. We introduced two deep learning architectures that provide promising accuracy with low and flexible computational complexity. We demonstrated their application to various digital constellations, and their ability to provide accurate soft posterior outputs. An important feature of one of our network is its ability to detect over multiple channel realizations with a single training.

Using neural networks as a general scheme in  MIMO detection still a long way to go and there are many open questions. These include their hardware complexity, robustness, and integration into full communication systems. 
Nonetheless, we believe this approach is  promising and has the potential to impact future communication systems. Neural networks can be trained on realistic channel models and tune their performance for specific environments. Their architectures and batch operation are more natural to hardware implementation than algorithms as SDR and SD. Finally, their multi-layer structure allows a flexible accuracy vs complexity nature as required by many modern applications. 


\section*{Acknowledgments}
We would like to thank Shai Shalev-Shwartz for many discussions throughout this research. In addition, we thank Amir Globerson and Yoav Wald for their ideas and help with the soft output networks. 

\ifCLASSOPTIONcaptionsoff
  \newpage
\fi

\bibliographystyle{IEEEtran}
\bibliography{main.bib}

\end{document}